\documentclass[11pt]{article}
\usepackage[dvips]{graphicx}
\usepackage{psfrag}
\usepackage{amssymb}
\usepackage{amsmath}
\usepackage{amscd}
\usepackage{latexsym}
\usepackage{bbm}

\catcode`@=11
\makeatletter\renewcommand{\section}{\@startsection
{section}{1}{\z@}{-3.5ex plus -1ex minus
    -.2ex}{2.3ex plus .2ex}{\large\bf }}

\makeatletter\renewcommand{\subsection}{\@startsection{subsection}{2}{\z@}{-3.25ex
plus -1ex minus
   -.2ex}{1.5ex plus .2ex}{\bf }}

\newcounter{saveeqn}

\catcode`@=12

\def\a{\alpha}
\def\b{\beta}
\def\ga{\gamma}
\def\la{\lambda}

\def\de{\delta}
\def\eps{\epsilon}
\def\ve{\varepsilon}

\def\th{\theta}

\def\vp{\varphi}

\newcommand{\C}{\mathbb C}
\newcommand{\R}{\mathbb R}

\newcommand{\Gcal}{{\cal G}}
\newcommand{\Acal}{{\cal A}}

\newcommand{\Mcal}{{\cal M}}
\newcommand{\Fcal}{{\cal F}}
\newcommand{\Ncal}{{\cal N}}

\newcommand{\gfrak}{{\mathfrak g}}

\def\1{{\bar 1}}

\def\diff{\textrm{d}}
\def\pa{\mbox{$\partial$}}
\def\sfrac#1#2{{\textstyle\frac{#1}{#2}}}
\def\+{\dagger}
\def\={\ =\ }
\def\und{\qquad\textrm{and}\qquad}
\def\and{\quad\textrm{and}\quad}
\def\with{\quad\textrm{with}\quad}
\def\for{\quad\textrm{for}\quad}

\def\Id{\mathrm{Id}}

\begin{document}

\begin{titlepage}
\setcounter{page}{0}
\begin{flushright}
ITP--UH--06/15
\end{flushright}

\hspace{2.0cm}

\begin{center}

{\LARGE\bf
Yang-Mills moduli space in the adiabatic limit
}

\vspace{12mm}

{\Large  Olaf Lechtenfeld \ and \ Alexander D.~Popov
}\\[8mm]

\noindent {\em
Institut f\"ur Theoretische Physik {\rm and} Riemann Center for Geometry and Physics\\
Leibniz Universit\"at Hannover \\
Appelstra\ss e 2, 30167 Hannover, Germany }\smallskip\\
{Email: lechtenf@itp.uni-hannover.de, popov@itp.uni-hannover.de}\\[6mm]

\vspace{10mm}

\begin{abstract}
\noindent
We consider the Yang-Mills equations for a matrix gauge group $G$ inside the future light cone of
4-dimensional Minkowski space, which can be viewed as a Lorentzian cone $C(H^3)$ over the 3-dimensional
hyperbolic space $H^3$. Using the conformal equivalence of $C(H^3)$ and the cylinder $\R\times H^3$,
we show that, in the adiabatic limit when the metric on $H^3$ is scaled down, classical Yang-Mills dynamics
is described by geodesic motion in the infinite-dimensional group manifold $C^\infty (S^2_\infty,G)$
of smooth maps from the boundary 2-sphere $S^2_\infty=\partial H^3$ into the gauge group $G$.
\end{abstract}

\end{center}
\end{titlepage}

\noindent
{\bf 1.}
Yang-Mills theory with Higgs fields governs three fundamental forces of Nature. It has a number of
particle-like solutions such as vortices, monopoles and instantons~\cite{1, 2, 3}. One may ask about the dynamics of
vortices and monopoles which evolve according to the second-order field equations of Yang-Mills-Higgs theory. In the
seminal paper~\cite{4} Manton suggested that in the ``slow-motion limit'' monopole dynamics can be described by
geodesics in the moduli space of static multi-monopole solutions.\footnote{
For nice reviews and a lot of references see e.g.~\cite{2,3,5}.}
This approach was extended both to vortices in $2{+}1$ dimensions (see e.g.~\cite{6} for a review)
and instantons in $4{+}1$ dimensions (see e.g.~\cite{7,8}).
In contrast, almost nothing is known about time-dependent solutions of pure Yang-Mills theory in $3{+}1$ dimensions.
Here we aim to partially fill this gap by applying Manton's approach to the Yang-Mills equations on Minkowski space.

\medskip

\noindent
{\bf 2}. We parametrize Minkowski space-time $\R^{3,1}$ with coordinates $x^\mu$, $\mu =0,1,2,3$, and the metric
\begin{equation}\label{1}
 \diff s^2 = \eta_{\mu\nu}\diff x^\mu \diff x^\nu \quad\with (\eta_{\mu\nu})=\mathrm{diag}(-1, 1, 1, 1)\ .
\end{equation}
In this article we fix an origin in $\R^{3,1}$ and consider the time evolution of Yang-Mills fields
in the interior of its light cone. For simplicity we will restrict ourselves to the future light cone~$L_+$
and its interior $T_+$ only, as the considerations for the past are similar. $L_+$ and $T_+$ are defined by
\begin{equation}\label{2}
 \bigl( \tau^2=0\ ,\ x^0>0 \bigr) \and \bigl( \tau^2>0\ ,\ x^0>0 \bigr)
 \quad\for \tau^2= -\eta_{\mu\nu} x^\mu x^\nu\ ,
\end{equation}
respectively.

On $T_+$ one can introduce global pseudospherical coordinates $(\tau , \chi , \theta ,\vp)$ by
\begin{equation}\label{3}
 x^0 = \tau\cosh\chi\ ,\quad
 x^1 = \tau\sinh\chi\sin\th\cos\vp\ ,\quad
 x^2 = \tau\sinh\chi\sin\th\sin\vp\ ,\quad
 x^3 = \tau\sinh\chi\cos\th
\end{equation}
and a range of
\begin{equation}
 \tau\in(0,\infty)\ ,\quad \chi\in[0,\infty)\ ,\quad \th\in[0,\pi]\ ,\quad \vp\in[0,2\pi)
\end{equation}
with the usual identifications and a harmless coordinate singularity at $\chi=0$.
The eigentime coordinate $\tau$ foliates $T_+$ into a family of hyperbolic 3-spaces $H^3(\tau)$
or `radius' $\tau$, each of which is built from spheres $S^2(\chi)$ of radius $\tau\sinh\chi$.
In these coordinates, the metric (\ref{1}) acquires the form
\begin{equation}\label{4}
 \diff s^2 \= - \diff\tau^2 + \tau^2\left\{\diff\chi^2 + \sinh^2\chi\, (\diff\th^2 + \sin^2\th\, \diff\vp^2)\right\}\ ,
\end{equation}
where the expression in the round brackets is the metric on  $S^2$
and the expression in the curly brackets is the metric on  $H^3$.
For any given $\tau$, the boundary $\partial H^3(\tau)$ is reached in the limit $\chi\to\infty$
and forms a 2-sphere $S^2_\infty$ `at infinity'.

The metric (\ref{4}) can be rewritten as
\begin{equation}\label{5}
 \diff s^2 \= - \diff\tau^2 + \tau^2\de_{ab}\,e^a\otimes e^b \=
 \tau^2\bigl(-(\tau^{-1}\diff\tau)^2 + \de_{ab}\,e^a\otimes e^b\bigr)\ ,
\end{equation}
where $\{e^a\}$ is a basis of one-forms on $H^3$ easily extracted from (\ref{4}).
{}From (\ref{5}) we recognize a cone over $H^3$, i.e.~$T_+=C(H^3)$,
which is conformally equivalent to a cylinder $\R\times H^3$ with the metric
\begin{equation}\label{6}
\diff s^2_{cyl} \= - \diff u^2 + \de_{ab}\,e^a\otimes e^b \quad\for u=\ln\tau
\end{equation}
and $H^3=H^3(\tau{=}1)$. We redenote the cylindrical coordinates,
\begin{equation}
 (u,\chi,\th,\vp) \= (y^0,y^1,y^2,y^3) \= (y^0,y^a) \quad\with a=1,2,3\ .
\end{equation}
From this point on we will work on the cylinder (\ref{6}) since Yang-Mills theory is conformally invariant.

\medskip

\noindent
{\bf 3.} We have set the stage to consider pure Yang-Mills theory on the cylinder $\R\times H^3$
with an arbitrary matrix gauge group $G$.
The Yang-Mills potential $\Acal =\Acal_{\mu}\diff y^\mu$ takes its value in the Lie algebra $\gfrak=\mathrm{Lie}\, G$
carrying a scalar product defined by the matrix trace Tr.
The field tensor $\Fcal =\diff\Acal + \Acal\wedge\Acal$ is defined as
\begin{equation}\label{7}
 \Fcal \=\sfrac12\Fcal_{\mu\nu}\,\diff y^\mu \wedge \diff y^\nu \quad\with\quad
 \Fcal_{\mu\nu} \=\partial_\mu\Acal_\nu - \partial_\nu\Acal_\mu + [\Acal_\mu , \Acal_\nu]\ ,
\end{equation}
and the Yang-Mills equations read
\begin{equation}\label{8}
 D_\mu\Fcal^{\mu\nu}\ :=\
 \sqrt{|\det g|}^{-1}\,\partial_\mu\bigl(\sqrt{|\det g|}\,\Fcal^{\mu\nu}\bigr) + [\Acal_\mu,\Fcal^{\mu\nu}]\=0\ ,
 \end{equation}
where $g=(g_{\mu\nu})$ is the metric (\ref{6}) on $\R\times H^3$.

For the metric (\ref{6}) we have
\begin{equation}\label{9}
 \Acal \= \Acal_0\diff y^0 + \Acal_a\diff y^a \=  \Acal_0\diff y^0 + \Acal_{H^3}\ ,
\end{equation}
\begin{equation}\label{10}
 \Fcal \= \Fcal_{0a}\,\diff y^0\wedge\diff y^a + \sfrac12\Fcal_{ab}\,\diff y^a\wedge\diff y^b \=
 \Fcal_{0a}\,\diff y^0\wedge\diff y^a + \Fcal_{H^3}\ .
\end{equation}
Employing the adiabatic approach \cite{4}, we deform the metric (\ref{6}) and introduce
\begin{equation}\label{11}
\diff s^2_{\ve} \= - \diff u^2 + \ve^2\de_{ab}\,e^a\otimes e^b\ ,
\end{equation}
where $\ve$ is a real positive parameter. Then $|\det g_{\ve}|=\ve^6 |\det g|$,
\begin{equation}\label{12}
 \Fcal^{0a}_\ve \= g^{00}_\ve  g^{ab}_\ve \Fcal_{0b} \=\ve^{-2}\Fcal^{0b} \und
 \Fcal^{ab}_\ve \=\ve^{-4}\Fcal^{ab}\ ,
\end{equation}
where in $\Fcal^{0a}$ and $\Fcal^{ab}$ the indices were raised by the non-deformed metric.

The adiabatic limit of scaling down the metric on $H^3$ is effected by the limit $\ve\to0$.
To avoid the $\ve^{-1}$ divergence of the Yang-Mills action functional, one has to impose
the vanishing of the curvature (\ref{10}) along $H^3$,
\begin{equation}\label{F0}
 \Fcal_{H^3}\=0\ ,
\end{equation}
which renders the connection $\Acal_{H^3}$ flat.
Substituting (\ref{12}) into the Yang-Mills equations on the cylinder $\R\times H^3$
with the metric (\ref{11}) and taking the adiabatic limit $\ve\to 0$ (corresponding to `slow $u$ evolution')
together with $\Fcal_{H^3}=0$, we obtain
\begin{equation}\label{13}
 g^{ab} D_a\Fcal_{b0}\=0\ ,
\end{equation}
\begin{equation}\label{14}
 D_0\Fcal_{0b}\=0\ ,
\end{equation}
which are, in fact, valid for any $\ve>0$ as well.

\medskip

\noindent {\bf 4.} Let us characterize the `static' Yang-Mills configurations, i.e.~the $u$-independent solutions
to~(\ref{F0}), following~\cite{Sal, Kori}. Any flat connection $\Acal_{H^3}$ on $H^3$ is formally pure gauge,
\begin{equation}\label{15}
 \Acal_{H^3}\=g^{-1}\hat\diff g \quad\with\quad \hat\diff =\diff y^a\sfrac{\partial}{\partial y^a}\ ,
\end{equation}
where $\hat\diff$ is the exterior derivative on $H^3$ and $g=g(y^a)$ is a smooth map from $H^3$ into the gauge group $G$.
Since $\partial H^3=S^2_\infty$ is not empty, the group of admissible gauge transformations is
\begin{equation}\label{16}
 \Gcal \= \left\{g\in C^\infty(H^3,G) \ \big|\ {g|_{\partial H^3}} = \Id\right\}\ .
\end{equation}
The boundary condition on $g$ obstructs the removal of
\begin{equation}\label{17}
 \Acal_{\pa H^3}\=g^{-1}\hat\diff g\,|_{S^2_\infty}
\end{equation}
by a gauge transformation and renders the flat connection (\ref{15}) non-trivial.
Hence, the solution space of the equation  $\Fcal_{H^3}=0$ is the infinite-dimensional group
\begin{equation}\label{Ncal}
 \Ncal \= C^\infty (H^3,G)\ ,
\end{equation}
and the moduli space is the quotient group
\begin{equation}\label{18}
 \Mcal \=  \Ncal / \,\Gcal \=  C^\infty (S^2_\infty, G)\ .
\end{equation}
The current groups (\ref{16}), (\ref{Ncal}) and (\ref{18}) as well as the corresponding moduli spaces 
of flat connections are well studied in the literature, see e.g.~\cite{Sal, Kori, Losev} and references therein. 
In fact, (\ref{16}) and (\ref{Ncal}) are groups of gauge transformations of bundles with and without framing 
over the boundary $S^2$ \cite{Don}, respectively, and the moduli space (\ref{18}) is their quotient. 
In our case, the framing (\ref{16}) is equivalent to imposing the Dirichlet boundary conditions, 
which are natural for Yang-Mills theory on manifolds with boundary~\cite{Don}.

Notice that the group (\ref{18}) contains as subgroups the loop group $C^{\infty}(S^1, G)$ as well as 
finite-dimensional submanifolds of finite-degree holomorphic maps from $\C P^1\cong S^2$ into K\"ahler coset 
spaces $G/H\subset G$. For a physical interpretation of the moduli space~(\ref{18}), 
we remark that the gauge equivalence classes of flat connections on~$H^3$ are neither solitonic nor instantonic 
in character, but rather describe different static Yang-Mills vacua. Here, the term `static' refers to 
our choice of Lorentz-invariant eigentime~$\tau$.\footnote{
For any adiabatic approximation one must pick some temporal foliation.} 
Since $\pi_2(G)=0$ for any compact connected finite-dimensional group $G$, 
the moduli space $\Mcal$ has just a single component.

\medskip

\noindent {\bf 5.} We introduce local coordinates $\phi^\a$ with $\a =1,2,\ldots$ on the moduli space
$\Mcal=C^\infty(S^2_\infty,G)$ and assume, following Manton, that $\Acal$ on the cylinder $\R\times H^3$ given by (\ref{9})
depends on $u$ (and hence on $\tau$) only via the moduli $\phi^\a (u)$. In other words, $\Acal_{H^3}=g^{-1}\hat\diff
g\bigl(\phi^\alpha(u);y^a\bigr)$,  $g\bigl(\phi^\a(u);\chi{\to}\infty\bigr)$ is determined by $\phi^\a(u)$ and
$\Acal_0(\phi^\a (u))$ will be fixed in a moment. This defines a map
\begin{equation}\label{19}
 \phi : \R\to\Mcal \quad\with \phi (u)=\{\phi^\a(u)\}\ .
\end{equation}
This map is not free -- it is constrained by (\ref{13}) and (\ref{14}). Since $\Acal_{H^3}$ belongs to
the solution space $\Ncal$ of flatness equations for any $u\in\R$,
its derivative $\partial_0\Acal_{H^3}$ is a solution of the flatness condition linearized around $\Acal_{H^3}$,
i.e.~$\partial_0\Acal_{H^3}$ belongs to the tangent space $T_\Acal \Ncal$.
With the help of the projection $\pi : \Ncal\to\Mcal$, one can decompose $\partial_0\Acal_a$ into two parts,
\begin{equation}\label{20}
 T_\Acal \Ncal \=\pi^* T_\Acal \Mcal \oplus T_\Acal \Gcal \qquad\Leftrightarrow\qquad
 \partial_0\Acal_a\=(\partial_0\phi^\a)\xi_{\a a} + D_a\eps_0\ ,
\end{equation}
where $\{\xi_\a =\xi_{\a a}\diff y^a\}$ is a local basis of vector fields on $\Mcal$, and
$\eps_0$ is a $\gfrak$-valued gauge parameter which is determined by the gauge-fixing equation
\begin{equation}\label{21}
 g^{ab} D_a\xi_{\a b}\=0 \qquad\Leftrightarrow\qquad g^{ab} D_a\partial_0\Acal_b\= g^{ab} D_aD_b\eps_0\ .
\end{equation}

Let us fix the gauge on $\R\times H^3$ by choosing $\Acal_0=\eps_0$.
Then (\ref{20})--(\ref{21}) imply that
\begin{equation}\label{22}
 \Fcal_{0b}\=\partial_0\Acal_b - D_b\Acal_0 \= \partial_0\Acal_b - D_b\eps_0 \=
 \dot\phi^\a\xi_{\a b}\=\pi_* \partial_0\Acal_b\ ,
\end{equation}
where the dot denotes the derivative with respect to $y^0=u$.
From (\ref{20})--(\ref{22}) we then see that (\ref{13}) is satisfied.
Furthermore, we obtain
\begin{equation}\label{23}
 \partial_0\Acal_a\=\dot\phi^\a\,\frac{ \partial\Acal_a}{ \partial\phi^\a} \qquad\Rightarrow\qquad
 \Acal_0\=\eps_0=\dot\phi^\a\,\eps_\a\ ,
\end{equation}
where the gauge parameters $\eps_\a$ can be found as solutions to
\begin{equation}\label{24}
 g^{ab} D_a D_b\eps_0\=g^{ab} D_a\frac{ \partial\Acal_b}{ \partial\phi^\a}\ .
\end{equation}

\medskip

\noindent
{\bf 6.} Substituting (\ref{22}) into the remaining equation (\ref{14}), we arrive at
\begin{equation}\label{25}
 g^{ab} \sfrac{\diff}{\diff u}(\dot\phi^\b\xi_{\b b})\= g^{ab}\dot\phi^\b[\xi_{\b b},\eps_0] \ .
\end{equation}
Let us multiply this equation with $\dot\phi^\a\xi_{\a a}$, apply Tr and integrate over $H^3$. This yields\footnote{
The right-hand side of (\ref{25}) disappears since
$g^{ab}\dot\phi^\a\dot\phi^\b\mathrm{Tr}\left([\xi_{\a a}, \xi_{\b b}]\eps_0\right)\equiv 0$.}
\begin{equation}\label{26}
 \sfrac{\diff}{\diff u}(G_{\a\b}\dot\phi^\a\dot\phi^\b)\=0\ ,
\end{equation}
where $G_{\a\b}$ are the metric components on the moduli space $\Mcal$, defined as
\begin{equation}\label{27}
 G_{\a\b}\= - \int_{H^3}\!\diff\,\mathrm{vol}\  g^{ab}\, \mathrm{Tr} (\xi_{\a a}\xi_{\b b})\ .
\end{equation}
This metric is the standard left-invariant metric on the Lie group. One can get it by left translations from the
Killing-Cartan metric on the tangent space at the identity in $\Mcal = C^{\infty}(S^2_{\infty}, G)$. 
However, its calculation may not be an easy task. We postpone its study to the future.

 Identifying $y^0=u$ with the length parameter on $\Mcal$,
i.e.~choosing the metric as
\begin{equation}
 \diff u^2\=G_{\a\b}\diff\phi^\a\diff\phi^\b\ ,
\end{equation}
(\ref{26}) becomes the geodesic equation on $\Mcal$ with affine parameter~$u$. To see them in more standard form, consider the
action
\begin{equation}\label{28}
 \tilde S \= \int\! \diff u\  \sqrt{G_{\a\b}\dot\phi^\a\dot\phi^\b}\ ,
\end{equation}
whose Euler-Lagrange equations are
\begin{equation}\label{29}
 \ddot\phi^\a + \Gamma^\a_{\b\ga} \dot\phi^\b \dot\phi^\ga - \dot\phi^\a\sfrac{\diff}{\diff u} \ln (G_{\b\ga}
 \dot\phi^\b\dot\phi^\ga)\=0\qquad\stackrel{\mathrm{(30)}}{\Longrightarrow}\qquad
 \ddot\phi^\a + \Gamma^\a_{\b\ga} \dot\phi^\b\dot\phi^\ga\=0\ ,
\end{equation}
where the Christoffel symbols are
\begin{equation}\label{30}
 \Gamma^\a_{\b\ga}\=\sfrac12\, G^{\a\la}\Bigl( \frac{\partial}{\partial \phi^\ga}\, G_{\b\la} +
 \frac{\partial}{\partial \phi^\b}\, G_{\ga\la} - \frac{\partial}{\partial \phi^\la}\, G_{\b\ga}\Bigr)\ .
\end{equation}
This derivation reflects the equivalence of the action (\ref{28}) and the functional
\begin{equation}\label{31}
 S\= \int\! \diff u\  G_{\a\b}\dot\phi^\a\dot\phi^\b\ .
\end{equation}
The latter is the effective Yang-Mills action in the adiabatic limit $\ve\to 0$ and stems from the term
\begin{equation}\label{32}
 \int_{\R\times H^3}\!\diff\,\mathrm{vol}\ \mathrm{Tr}(\Fcal_{0a}\Fcal^{0a})
\end{equation}
in the original Yang-Mills action functional. Since $\Mcal$ is a Lie group, one can construct geodesics as one-parameter
subgroups, and we intend to do this in a separate publication. The physical meaning of the moduli parameters $\phi^\a$ will
become clear from the properties of such solutions.

If we {\it assume\/} that $\Fcal_{H^3}=0$  for any $\tau= e^u$, then (\ref{13})--(\ref{14}) form all Yang-Mills
equations on $\R\times H^3$ for any $\ve\ne 0$ including $\ve = 1$.\footnote{
In general  $\Fcal_{H^3}=0$  is mandatory unless $\ve\to 0$.}
Their solutions
\begin{equation}
 (\Acal_0, \Acal_a)\=\bigl(\dot\phi^\a\eps_\a\ ,\ g^{-1}\partial_a g(\phi^\a;\chi,\th,\vp)\bigr)
 \quad\with \phi=\phi(u)
\end{equation}
carry electrical but no magnetic charge since $\Fcal_{0a}{\ne}0$ while $\Fcal_{ab}{=}0$. From the implicit function theorem it
follows that for any solution $\Acal^{\ve=0}_\mu$ defined by $\phi$ satisfying (\ref{29}) there exist nearby solutions
$\Acal^{\ve>0}_\mu$ of the Yang-Mills equations for $\ve$ sufficiently small, and we conjecture that the moduli space of all
geodesics (\ref{29}) in $C^\infty (S^2_\infty, G)$ is bijective to the moduli space of solutions to the Yang-Mills equations.

\medskip

\noindent {\bf 7.} In conclusion, we reduced Yang-Mills theory on Minkowski space in a certain adiabatic limit to a
one-dimensional sigma model with the target space $\Mcal = C^\infty (S^2_\infty, G)$, which should capture the low-energy
dynamics of the gauge theory. We note that the group $C^{\infty}(\Sigma,G)$ of smooth maps from a Riemannian surface $\Sigma$
(including the case of $S^2$) into a Lie group $G$ has been considered by mathematicians (see e.g.~\cite{9, 10}) but did not
yet find a true application in physics. This short article indicates relations of such groups with Yang-Mills theory in four
dimensions.

\medskip

\noindent
{\bf Acknowledgements} \\[2pt]
This work was partially supported by the Deutsche Forschungsgemeinschaft grant LE 838/13.

\end{document}